\documentclass[10pt,conference]{IEEEtran}
\usepackage{cite}
\usepackage{graphicx}
\usepackage{epsfig}
\usepackage{psfrag}
\usepackage{subfigure}
\usepackage{url}
\usepackage{stfloats}
\usepackage{amsmath}
\usepackage{color}
\usepackage{amssymb}
\usepackage{epsfig}
\usepackage{array}

\newtheorem{theorem}{Theorem}

\usepackage{xspace}
\usepackage{bbm}
\usepackage{mathrsfs}
\newcommand{\Tc}{\mathcal{T}}

\newcommand{\supp}{{\textrm{supp}}}
\newcommand{\Xv}{{\bf X}}

\newcommand{\Yv}{{\bf Y}}

\newcommand{\Vv}{{\bf V}}
\newcommand{\Sv}{{\bf S}}

\let\P\relax
\DeclareMathOperator\P{\sf P}
\def\textiid{i.i.d.\@\xspace}
\newcommand\iid{\ifmmode\text{ i.i.d. } \else \textiid \fi}

\begin{document}

\title{Sparse Signal Recovery in the Presence of Intra-Vector and Inter-Vector Correlation}

\author{
\authorblockN{Bhaskar D. Rao}
\authorblockA{Dept.\ of Electrical and Computer Eng.\\
University of California, San Diego\\
La Jolla, 92093-0407, USA\\
Email: brao@ucsd.edu} \and
\authorblockN{Zhilin Zhang}
\authorblockA{Dept.\ of Electrical and Computer Eng.\\
University of California, San Diego\\
La Jolla, 92093-0407, USA\\
Email: z4zhang@ucsd.edu}
\and
\authorblockN{Yuzhe Jin}
\authorblockA{Microsoft Research\\
One Microsoft Way\\
Redmond, WA 98052, USA\\
Email: yuzjin@microsoft.com}
}

% avoiding spaces at the end of the author lines is not a problem with
% conference papers because we don't use \thanks or \IEEEmembership
\maketitle

\begin{abstract}
This work discusses the problem of sparse signal recovery when there is correlation among the values of non-zero entries. We examine intra-vector correlation in the context of the block sparse model and inter-vector correlation in the context of the multiple measurement vector model, as well as their combination. Algorithms based on the sparse Bayesian learning are presented and the benefits of incorporating correlation at the algorithm level are discussed. The impact of correlation on the limits of support recovery is also discussed highlighting the different impact intra-vector and inter-vector correlations have on such limits.
\end{abstract}

%========================================================================================
\section{Introduction}
\label{sec:intro}
%========================================================================================

The problem of sparse signal recovery has many potential applications \cite{brao:icassp98,elad2010sparse} and has received much attention in recent years
with the development of compressed sensing (CS) \cite{donoho2006compressed,candes2006near}. The general Multiple Measurement Vector (MMV) model is given by \cite{Cotter2005}
\begin{eqnarray}
\mathbf{Y}= \mathbf{\Phi} \mathbf{X} + \mathbf{V}.
\label{equ:MMV basicmodel}
\end{eqnarray}
Here $\mathbf{Y} \triangleq [\mathbf{Y}_{\cdot 1},\cdots,\mathbf{Y}_{\cdot L}] \in \mathbb{R}^{N \times L}$ is an available measurement matrix consisting of $L$ measurement vectors. $\mathbf{\Phi} \in \mathbb{R}^{N \times M} (N \ll M)$ is a known matrix, and any $N$ columns of $\mathbf{\Phi}$ are linearly independent. $\mathbf{X}\triangleq [\mathbf{X}_{\cdot 1},\cdots,\mathbf{X}_{\cdot L}] \in \mathbb{R}^{M \times L}$ is an unknown and full column-rank matrix
of interest. A key assumption here is that $\mathbf{X}$ has only a few non-zero rows. $\mathbf{V}$ is a noise matrix. The special case of $L = 1$ is the widely studied Single Measurement Vector (SMV) problem in CS and in this context we use $\mathbf{x}$ to denote the vector of interest.

\section{Structure in $\mathbf{X}$}
In the basic SMV and MMV models no additional assumptions are usually made. However, in many applications additional structure
on $\mathbf{X}$ is available and we now discuss a few of them.

(1) For the SMV problem, in contrast to the usual assumptions that the locations
of non-zero entries are independently and uniformly distributed, some dependency in the locations is assumed \cite{faktor2010exploiting,ModelCS,Cevher2010}.
Incorporating this structure is important from an application point of view and this structure can
be exploited to improve the performance of algorithms.

(2) In the SMV problem a widely studied structure is block/group structure \cite{grouplasso,Eldar2009}. With this structure, $\mathbf{x}$ can be viewed as
a concatenation of $g$ blocks, i.e.
\begin{eqnarray}
\mathbf{x} = [ \underbrace{x_1,\cdots,x_{d_1}}_{\mathbf{x}_1^T},   \cdots,  \underbrace{x_{d_{g-1}+1},\cdots,x_{d_g}}_{\mathbf{x}_g^T}]^T
\label{equ:partition}
\end{eqnarray}
where $d_i (\forall i)$ are not necessarily the same. Among the $g$ blocks, only $k$ blocks are nonzero, where $k \ll g$.
This can be viewed as a special case of modeling the distribution of the locations of the non-zero entries, but is worthy of special attention because of its application potential. In general, no additional assumption is made about the entries in each nonzero block. Motivated by applications, it appears reasonable to
assume that the entries in each non-zero block are correlated \cite{zhang2012TSP,zhang2012TBME}. We refer to this as intra-block correlation and will discuss it in detail in Section
\ref{subsec:intra}.

(3) In the basic MMV problem, the typical assumption made is that the vectors in $\mathbf{X}$ share a common sparsity profile. This leads to
non-zero rows in $\mathbf{X}$. One can impose additional structure. One possibility could be dependency in the locations of the non-zero rows. And the other is correlation between the entries in each of the non-zero rows \cite{zhang2011IEEE,ziniel2011efficient}. We refer to the correlation as inter-vector correlation and will discuss it in Section \ref{subsec:inter}.

(4) One can combine the above-mentioned two types of structure and consider the problem of block sparsity in the MMV problem. This leads to
the consideration of correlated non-zero blocks of rows in $\mathbf{X}$. The challenge in this context is efficiently modeling and estimating
the correlation structure.

(5) The time-varying sparsity model is a natural extension of the MMV model \cite{Vaswani10,ziniel2010tracking,Cichocki2008}. It considers the case when the support of each column of $\mathbf{X}$ is time-varying. The time-varying structure calls for modeling both the variation in the number and locations of the non-zero entries as well as the correlation of the non-zero entries.

%==========================================================================================
\section{Intra-Vector and Inter-Vector Correlation}

\subsection{Intra-Vector Correlation}
\label{subsec:intra}

For the SMV problem with the block structure (\ref{equ:partition}), a number of algorithms have been proposed, such as the Group Lasso \cite{grouplasso}. But few  consider correlation within each block $\mathbf{x}_i(\forall i)$, namely the intra-block correlation.

To exploit the intra-block correlation, we have proposed the the block sparse Bayesian learning (bSBL) framework \cite{zhang2012TSP}, an extension of the basic SBL framework \cite{Tipping2001}. We review it in the following.

In this framework, each block $\mathbf{x}_i \in \mathbb{R}^{d_i \times 1}$ is assumed to satisfy a parameterized multivariate Gaussian distribution:
\begin{eqnarray}
p(\mathbf{x}_i) \sim  \mathcal{N}(\textbf{0},\gamma_i \mathbf{B}_i), \quad \forall i
\end{eqnarray}
Here $\gamma_i$ is a nonnegative parameter controlling the block-sparsity of $\mathbf{x}$. When $\gamma_i=0$, the $i$-th block becomes zero. During the learning procedure most $\gamma_i $ tend to be zero, due to the mechanism of automatic relevance determination \cite{Tipping2001}. Thus, sparsity in the block level is encouraged. $\mathbf{B}_i \in \mathbb{R}^{d_i \times d_i}$ is a positive definite matrix, capturing the correlation structure within the $i$-th block $\mathbf{x}_i$. The prior of $\mathbf{x}$ is $p(\mathbf{x}) \sim  \mathcal{N}(\textbf{0},\mathbf{\Sigma}_0)$, where $\mathbf{\Sigma}_0$ is a block-diagonal matrix with each principal block given by $\gamma_i \mathbf{B}_i$. Assume the noise vector satisfies $p(\mathbf{v}) \sim  \mathcal{N}(\textbf{0},\lambda \mathbf{I})$, where $\lambda$ is a positive scalar. Therefore, the posterior of $\mathbf{x}$ is given by $p(\mathbf{x}|\mathbf{y}; \lambda, \{\gamma_i, \mathbf{B}_i\}_{i=1}^g) = \mathcal{N}(\boldsymbol{\mu}_x, \mathbf{\Sigma}_x)$ with $\boldsymbol{\mu}_x  = \mathbf{\Sigma}_0 \mathbf{\Phi}^T \big(\lambda \mathbf{I} +  \mathbf{\Phi} \mathbf{\Sigma}_0 \mathbf{\Phi}^T \big)^{-1} \mathbf{y}$ and $\mathbf{\Sigma}_x = (\mathbf{\Sigma}_0^{-1} + \frac{1}{\lambda}\mathbf{\Phi}^T \mathbf{\Phi})^{-1}$. Once the hyperparameters $\lambda, \{\gamma_i, \mathbf{B}_i\}_{i=1}^g$ are estimated, the Maximum-A-Posterior (MAP) estimate of $\mathbf{x}$, denoted by $\widehat{\mathbf{x}}$, can be directly obtained from the mean of the posterior, i.e.
\begin{eqnarray}
\widehat{\mathbf{x}} \leftarrow \boldsymbol{\mu}_x  = \mathbf{\Sigma}_0 \mathbf{\Phi}^T \big(\lambda \mathbf{I} +  \mathbf{\Phi} \mathbf{\Sigma}_0 \mathbf{\Phi}^T \big)^{-1} \mathbf{y}.
\label{equ:bsbl_x}
\end{eqnarray}

The hyperparameters are generally estimated by a Type II maximum likelihood procedure \cite{Tipping2001}. This is equivalent to minimizing the following negative log-likelihood \cite{zhang2012TSP} with respect to each hyperparameter
\begin{eqnarray}
\mathcal{L}(\lambda, \{\gamma_i, \mathbf{B}_i\}_{i=1}^g) &\triangleq & \log| \mathbf{\Sigma}_y  |  + \mathbf{y}^T \mathbf{\Sigma}_y^{-1} \mathbf{y},
\label{equ:bsbl_costfunc}
\end{eqnarray}
where $\mathbf{\Sigma}_y \triangleq  \lambda \mathbf{I} + \mathbf{\Phi} \mathbf{\Sigma}_0 \mathbf{\Phi}^T$. A number of optimization approaches are available for estimating the hyperparameters \cite{zhang2012TSP}. Here we only present the results using the Expectation-Maximization (EM) method:
\begin{eqnarray}
\gamma_i  &\leftarrow &  \frac{1}{d_i} \mathrm{Tr}\big[ \mathbf{B}_i^{-1} \big( \mathbf{\Sigma}_x^i + \boldsymbol{\mu}_x^i (\boldsymbol{\mu}_x^i)^T\big)\big],\quad \forall i \label{equ:bsbl_gamma}\\
\lambda  & \leftarrow & \frac{\|\mathbf{y}-\mathbf{\Phi} \boldsymbol{\mu}_x\|_2^2 + \sum_i \mathrm{Tr}( \mathbf{\Sigma}_x^i (\mathbf{\Phi}^i)^T \mathbf{\Phi}^i)}{M} \label{equ:bsbl_lambda} \\
\mathbf{B}_i & \leftarrow & \mathrm{Toeplitz}([1,\overline{r},\cdots,\overline{r}^{d_i-1}]), \quad \forall i \label{equ:bsbl_B}
\end{eqnarray}
where  $\boldsymbol{\mu}_x^i \in \mathbb{R}^{d_i \times 1}$ is the corresponding $i$-th block in $\boldsymbol{\mu}_x$, and $\mathbf{\Sigma}_x^{i} \in \mathbb{R}^{d_i \times d_i}$ is the corresponding $i$-th principal diagonal block in $\mathbf{\Sigma}_x$. In (\ref{equ:bsbl_B}),  $\overline{r} \triangleq \mathrm{sign}(\frac{\overline{m}_1}{\overline{m}_0}) \min\{|\frac{\overline{m}_1}{\overline{m}_0}|,0.99\}$, where $\overline{m}_0 \triangleq \sum_{i=1}^g m_0^i$ and $\overline{m}_1 \triangleq \sum_{i=1}^g m_1^i$. Here $m_0^i$ and $m_1^i$ are the averages of the entries along the main diagonal and the main sub-diagonal of $\overline{\mathbf{B}}_i$, which is learned by the rule: $ \overline{\mathbf{B}}_i \leftarrow \frac{1}{\gamma_i} \big[ \mathbf{\Sigma}_x^i + \boldsymbol{\mu}_x^i (\boldsymbol{\mu}_x^i)^T \big]$. The resulting algorithm, denoted by BSBL-EM, then iterates over (\ref{equ:bsbl_x}) (\ref{equ:bsbl_gamma}) (\ref{equ:bsbl_lambda}) (\ref{equ:bsbl_B}) until convergence.

Extensive experiments have shown that the algorithms derived from the bSBL framework have the best recovery performance among existing algorithms \cite{zhang2012TSP} and shed light on various aspects of the intra-block correlation structure, including benefits of exploiting the correlation, guidance
on how to modify existing algorithms to exploit the correlation \cite{zhang2011ICML}, modification to deal with block sparsity with unknown block partition \cite{zhang2012TSP}, and applications to problems with less sparsity \cite{zhang2012TBME}.

\subsection{Inter-Vector Correlation}
\label{subsec:inter}

This is the situation in the MMV model (\ref{equ:MMV basicmodel}) where there is correlation among the entries in each non-zero row of $\mathbf{X}$. To deal with this situation, we assume the rows $\mathbf{X}_{i\cdot}$ ($\forall i$) are mutually independent, and the density distribution of each $\mathbf{X}_{i\cdot}$ is parameterized multivariate Gaussian, given by
\begin{eqnarray}
p(\mathbf{X}_{i\cdot};\gamma_i,\mathbf{B}_i) \sim \mathcal{N}(\mathbf{0},\gamma_i \mathbf{B}_i), \quad  i=1,\cdots,M
\nonumber
\end{eqnarray}
where $\gamma_i$ is a nonnegative hyperparameter controlling the row sparsity of $\mathbf{X}$. When $\gamma_i = 0$,  the associated $\mathbf{X}_{i\cdot}$ becomes zero. $\mathbf{B}_i$ is a positive definite  matrix that captures the  correlation structure of $\mathbf{X}_{i\cdot}$. Note that by letting $\mathbf{y}=\mathrm{vec}(\mathbf{Y}^T)$, $\mathbf{D} = \mathbf{\Phi} \otimes \mathbf{I}_L$, $\mathbf{x}=\mathrm{vec}(\mathbf{X}^T)$, and $\mathbf{v}=\mathrm{vec}(\mathbf{V}^T)$, we can transform the MMV model to the following SMV model \cite{zhang2011IEEE,Eldar2009}
\begin{eqnarray}
\mathbf{y}= \mathbf{D} \mathbf{x} + \mathbf{v}, \nonumber
\end{eqnarray}
where $\mathbf{x}$ has the block partition (\ref{equ:partition}) with $d_i=L(\forall i)$. Therefore, all the algorithms derived from the bSBL framework \cite{zhang2012TSP} can be applied to the MMV model. For more details, the reader is referred to \cite{zhang2011IEEE,Jing2012CVPR}. For convenience, the resulting algorithms are together called the T-SBL family. Interestingly,
the role of the correlation structure on the performance of existing MMV algorithms is found to be quite different from that
of intra-vector correlation \cite{zhang2012TSP}. Some explanation to this observation is provided in Section \ref{sec:theory}. As in the inter-vector case, algorithms in the T-SBL family provide insight into how to modify existing MMV algorithms that operate
in the $\mathbf{X}$-space to incorporate inter-vector correlation \cite{zhang2011ICML,Jing2012CVPR}.

In some applications the matrix $\mathbf{X}$ has both the intra-vector correlation and the inter-vector correlation. This correlation structure can be exploited as well by extending the bSBL framework. Assume $\mathbf{X}$ can be partitioned into a number of blocks, and the $i$-th block consists of $d_i$ rows. Then a key issue is how to model the correlation structure in each block. The most general
model would involve stacking the rows of a block and using a $d_i L \times d_i L$ matrix to model the correlation in this block. But estimating such a model from a small number of measurement vectors can lead to overfitting and unreliable estimates. Thus, simplified models are needed, and in this context the Kronecker model has support from applications. The overall correlation structure in the $i$-th block is modeled as $\mathbf{R}^i = \mathbf{R}_t^i \otimes \mathbf{R}_s^i,$ where $\mathbf{R}_t^i$ captures the
inter-vector correlation in this block and $\mathbf{R}_s^i$ captures the intra-vector correlation. Understanding the role of the correlation and how accurately to model and incorporate correlation is an interesting topic for future study.

\subsection{Time-Varying Sparsity Model}
\label{subsec:vary}

The time-varying sparsity model is a natural extension of the MMV model.
It considers the case when the support of each column of $\mathbf{X}$ is time-varying.
The transition from the stationary models, assumed so far, to the non-stationary situation
opens up an abundance of options akin to past work on tracking which has led to
adaptive filters, Kalman Filters and so on.

The measurement model in this case is given by
\begin{eqnarray}
\mathbf{y}_t= \mathbf{\Phi} \mathbf{x}_t + \mathbf{v}_t, \; t = 0,1,2,...
\label{equ:TV model}
\end{eqnarray}
Here, $\mathbf{y}_t \in \mathbb{R}^{N \times 1}$ is a measurement vector, $\mathbf{x}_t\in \mathbb{R}^{M \times 1}$ is the sparse signal with time-varying sparsity, and $\mathbf{v}_t$ is a noise vector.

A model for generating signals $\mathbf{x}_t$ with time-varying sparsity is needed both for developing optimal algorithms
and for systematic evaluation of algorithms developed. Drawing inspiration from applications like neuroelectromagnetic source localization, the measurement data can be viewed as being generated by a sequence of
events leading to an approximate piecewise stationary model.
Each stationary segment leads to an MMV model, which involves a sparsity pattern and a multivariate time series for the
nonzero entries that lasts a certain duration. The time series maybe modeled as a multivariate random signal with certain statistical properties
or a deterministic model. For the statistical case, one can use a multivariate AR process to model the signal. For the deterministic
case, one can assume it is the response of a dynamical system to an impulse input, e.g. a set of second order difference equations.

The transition from event to event may be completely random or structured. Completely random means the sparsity pattern
changes in an independent manner and the number of non-zero entries at a given time always lies in a given range. Structured
means that the sparsity change is more gradual, i.e. few entries get turned off and a new set of small entries are turned on
potentially in an asynchronous manner. A model with such reasonable flexibility will be very useful for generation of data and
testing of algorithms.

To deal with time-varying sparsity, several algorithms have been proposed, such as SOB-M-FOCUSS \cite{Cichocki2008}, message passing algorithms \cite{ziniel2010tracking}, and Least-Square Compressed Sensing (LS-CS) \cite{Vaswani10}. Since the support of $\mathbf{x}_t$ is changing slowly,  we can view such a time-varying  sparsity model as a concatenation of several MMV models \cite{zhang2011ICML}, where in each MMV model the support does not change. Therefore, algorithms in the T-SBL family can be used in this model. Note that here exploiting the multiple measurement vectors is important because of the enhanced support-recovery ability afforded by the MMV model as discussed in Section \ref{sec:theory}. And we will illustrate this benefit in Section \ref{sec:experiment}.

\section{Limits of Support Recovery}
\label{sec:theory}
An interesting question is the limits of sparse signal recovery algorithms, i.e., under what conditions is any algorithm capable of recovering the locations of the non-zero entries. Such results can potentially be also useful in understanding the role of the correlation structure in the support recovery task. Previous literature discussing the performance limits of sparse signal recovery can be divided into two categories. The first category of analysis focuses on the performance of practical algorithms \cite{donoho2006compressed,Tao05a,Tropp04,ZhaoYu2006,Bresler_ICIP_1996,EldarRauhut,OWainJ}. The second category of performance analysis focuses on the performance limits of the theoretical algorithms with combinatorial complexity \cite{Wain09b,Fletcher09IT,Akcakaya,JKR10}. In this paper, we consider the information theoretic performance limit of support recovery that governs any algorithm, which belongs to the second category as described above.

Let $W$ denote a matrix with all elements being non-zero. Define the generative model for the sparse signal $\Xv$ as
\begin{align}
X_{s,i} = \left\{ \begin{array}{ll}
w_{j,i} & \mbox{if $s=S_j$},\\
0 & \mbox{if $s\notin \{S_1, ..., S_K\}$}.\end{array} \right.
\label{signal_model}
\end{align}
The support of $\Xv$, denoted by $\supp(\Xv)$, is the set of indices corresponding to the non-zero rows of $\Xv$, i.e., $\supp(\Xv) = \{S_1,...,S_K\}$. According to the signal model (\ref{signal_model}), $|\supp(\Xv)|= K$. We assume $K$ is known.

Upon observing the noisy measurement $\Yv$, the goal is to recover the indices of the non-zero rows of $\Xv$. A support recovery map is defined as
\begin{align}
d: \mathbb{R}^{N\times L} \longmapsto 2^{[M]}.
\label{recon_algorithm}
\end{align}
We further define the average probability of error by
\[
\P\{d(\Yv)\neq \textmd{supp}(\Xv(W, \Sv))\}
\]
for each (unknown) signal value matrix $W\in\mathbb{R}^{K\times L}$. Note that the probability is averaged over the randomness of the locations of the non-zero rows $\Sv$, the measurement matrix $\bf{\Phi}$, and the measurement noise $\Vv$.

We consider the support recovery of a sequence of sparse signals generated with the same signal value matrix $W$. In particular, we assume that $K$ and $L$ are fixed. Define the auxiliary quantity
\begin{align}
c(W)\triangleq \min_{\Tc\subseteq [K]}\left[\frac{1}{2|\Tc|}\log\det\left(I + \frac{\sigma_{\phi}^2}{\sigma_v^2}\underline{W}_{\Tc}^\intercal \underline{W}_{\Tc}\right)  \right],
\end{align}
where $\underline{W}_{\Tc}$ denotes a matrix formed by  appropriately choosing a set of rows indexed by $\Tc$ from $W.$
The following two theorems summarize the performance limits in support recovery of sparse signals. The notation $N_M$ implies the possible dependency between $N$ and $M$.
\begin{theorem}
\label{theorem1}
If
\begin{align}
\limsup_{M\rightarrow\infty}\frac{\log M}{N_M} < c(W)
\end{align}
then there exists a sequence of support recovery maps $\{d^{(M)}\}_{M=K}^\infty,d^{(M)}:\mathbb{R}^{N_M\times L}\mapsto 2^{[M]}$, such that
\begin{align}
\lim_{M\rightarrow \infty}\P\{d(\Yv)\neq \textmd{supp}(\Xv(W, \Sv))\} =0.
\end{align}
\end{theorem}

\begin{theorem}
If
\begin{align}
\limsup_{M\rightarrow\infty}\frac{\log M}{N_M} > c(W)
\end{align}
then for any sequence of support recovery maps $\{d^{(M)}\}_{M=K}^\infty,d^{(M)}:\mathbb{R}^{N_M\times L}\mapsto 2^{[M]}$, \begin{align}
\liminf_{M\rightarrow \infty}\P\{d(\Yv)\neq \textmd{supp}(\Xv(W, \Sv))\} >0.
\end{align}
\end{theorem}

Theorems~1 and 2 together indicate that $N = \frac{1}{c(W)\pm \epsilon}\log M$ is the sufficient and necessary number of measurements per measurement vector to ensure asymptotically successful support recovery. The constant $c(W)$ explicitly captures the role of the non-zero entries in the performance tradeoff.

To understand the result and its implication, we need to examine the structure of the non-zero matrix $W$. Assume $L < K$, then the quantity $c(W)$, with mild assumptions on the non-zero entries, grows linearly with $L$ \cite{Paulraj}. This fact indicates that support recovery in the MMV problem greatly benefits from the presence of new measurements. Meanwhile, Theorems~1 and 2 characterize the role of each non-zero entry in the matrix $\Xv$ in the performance limit of support recovery of sparse signals. Indeed, adding different measurement vectors may cause drastically different performance gains. As a special case, when the columns of the non-zero signal matrix $W$ are identical, the performance gain of having MMV is equivalent to merely reducing the noise level by a factor of $L$. However, by properly constructing a matrix $W$ with certain rank conditions imposed on its submatrices, the performance limit of support recovery can enjoy a much larger gain as a result of, in the language of MIMO wireless communication, a multiplexing gain. For the SMV block sparsity model where $L=1$, no such benefit accrues. However, the norm of the blocks contributes to a signal-to-noise ratio gain. It is useful to note the analysis so far is conducted with a fixed $W$. For random non-zero entries, one can use the results in the two theorems above as the instantaneous capacity and conduct an outage analysis \cite{Paulraj,JKR10}. In the context of random entries, the blocks, under mild assumptions, provide a diversity gain that greatly improves the performance of block sparsity algorithms with known block size \cite{Eddy_2012}.

%===================================================================================
\section{Experiments}
\label{sec:experiment}
%===================================================================================

Three representative experiments were performed. Each experiment is based on 500 trials. In each trial the matrix  $\mathbf{\Phi} \in \mathbb{R}^{N \times M}$ was generated to be a Gaussian random matrix with columns to be unit norm. We chose the MSE as a performance index in noisy experiments, and the Success Rate as a performance index in noiseless experiments. The success rate was defined as the ratio of the number of successful trials to the number of total trials, while a successful trial was defined as the one when $\mathrm{MSE} \leq 10^{-6}$.

\textbf{Experiment 1: Effect of Intra-Block Correlation in a SMV Model}. In this noiseless experiment we studied the effect of intra-block correlation with the use of the BSBL-EM algorithm presented in Section \ref{subsec:intra}. The matrix $\mathbf{\Phi}$ was of the size $100 \times 300$. The sparse signal $\mathbf{x}$ consisted of 75 blocks with identical size. Only 20 of the blocks were non-zero. Entries in every non-zero block were modeled as an AR(1) process with the same AR coefficient $\beta$. $\beta$ assumed values ranging from -0.99 to 0.99. The experiment was then repeated for each value of $\beta$. BSBL-EM was performed in two ways, namely adaptively learning the intra-block correlation and completely ignoring the correlation (i.e. set $\mathbf{B}_i = \mathbf{I}(\forall i)$).

\begin{figure}[t]
\begin{minipage}[b]{.48\linewidth}
  \centering
  \centerline{\epsfig{figure=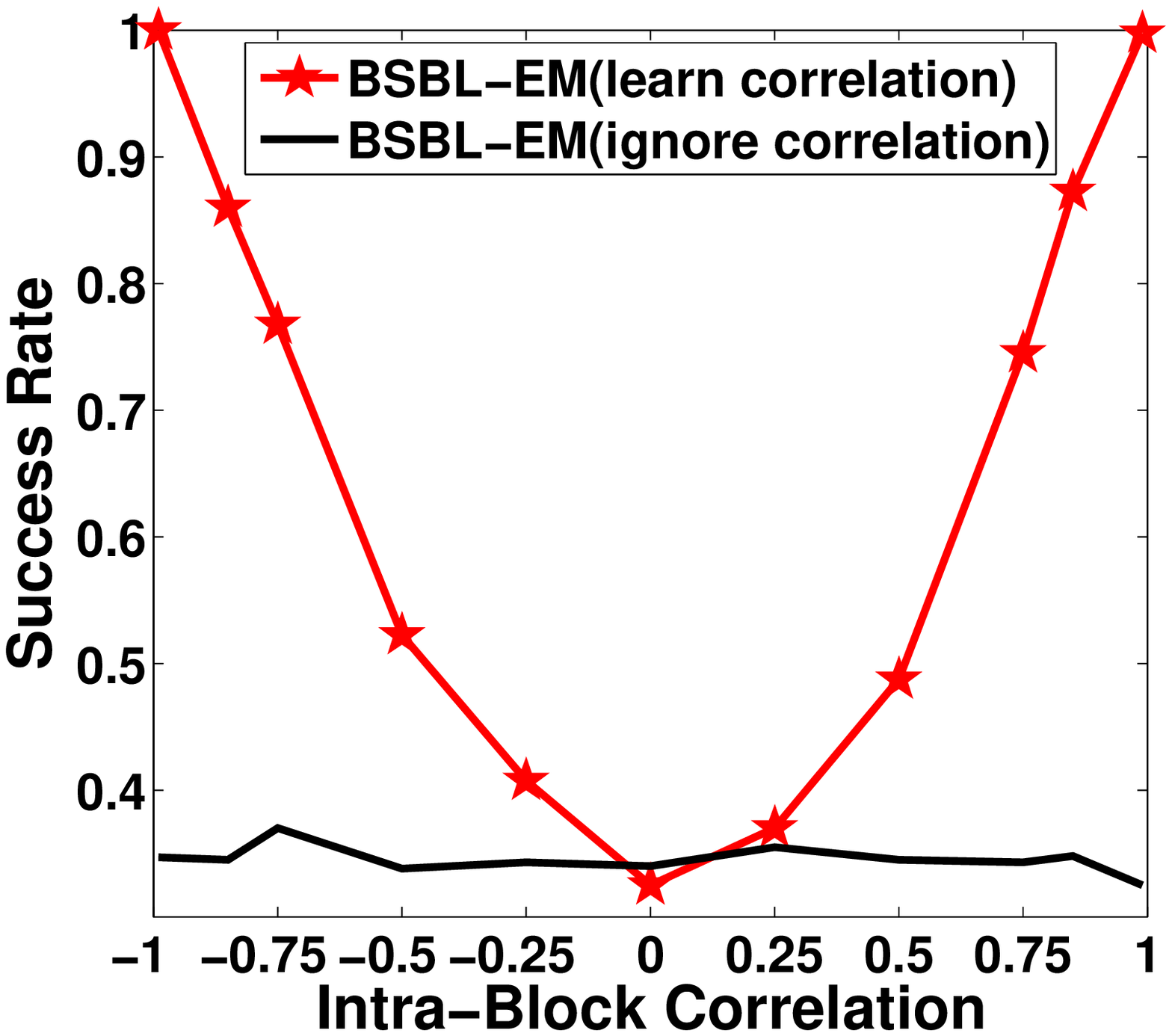,width=4.5cm,height=3.5cm}}
  \centerline{\footnotesize{(a)}}
\end{minipage}
\hfill
\begin{minipage}[b]{0.48\linewidth}
  \centering
  \centerline{\epsfig{figure=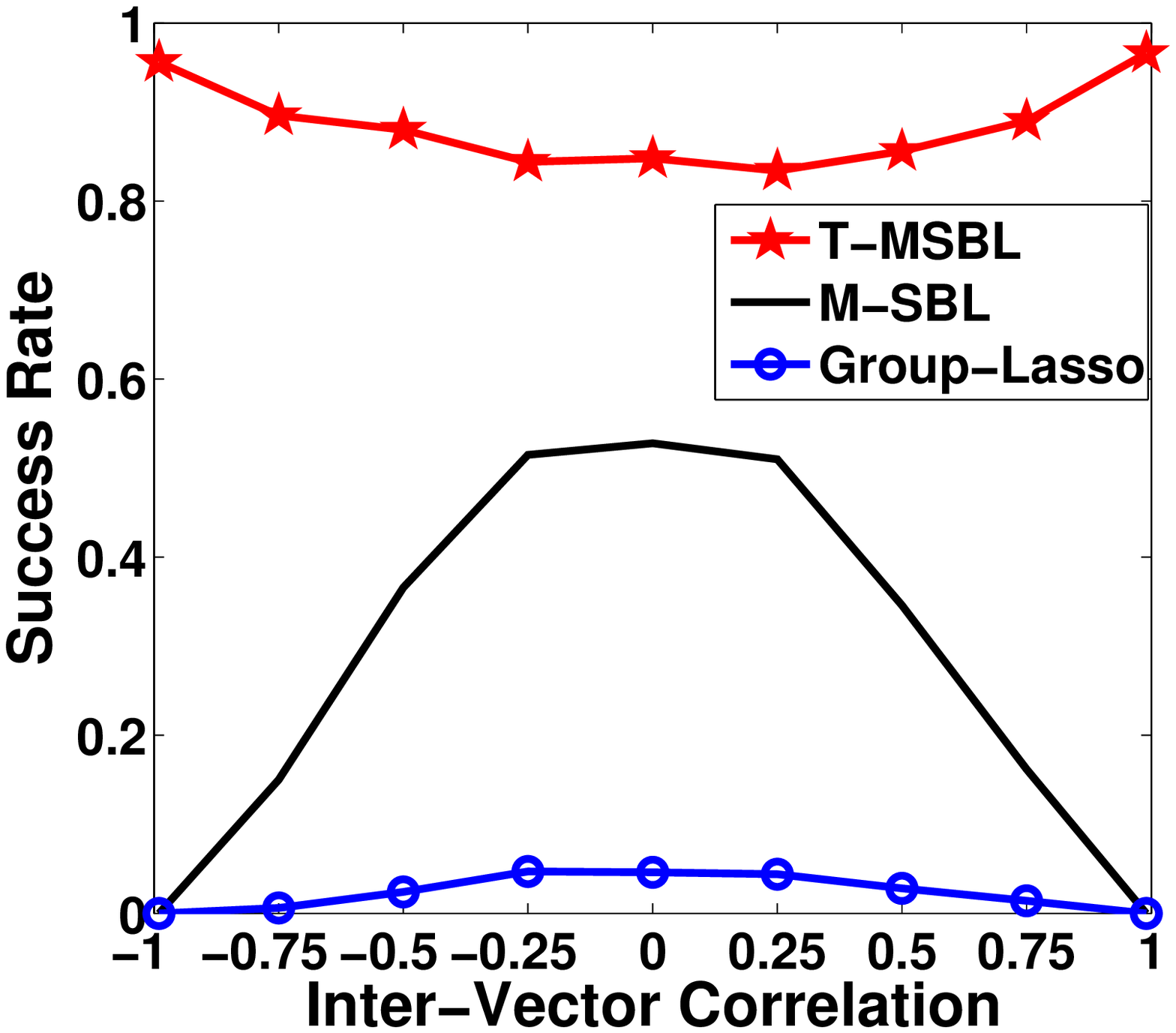,width=4.5cm,height=3.5cm}}
  \centerline{\footnotesize{(b)}}
\end{minipage}
\caption{(a) Effects of intra-block correlation on algorithm performance. (b) Effects of inter-vector correlation on algorithm performance. }
\label{fig:Simulation12}
\end{figure}

The result (Fig.\ref{fig:Simulation12} (a)) clearly shows that when correlation is exploited, BSBL-EM has improved performance  with the increase in the correlation. However, when the correlation is not exploited,
the performance is unchanged with correlation. Note that the latter phenomenon was also observed from existing  algorithms which do not exploit the correlation \cite{zhang2012TSP}.

\textbf{Experiment 2: Effect of Inter-Vector Correlation in an MMV model}. Next we studied the effect of inter-vector correlation in a noiseless MMV experiment, where $N=25, M=125, L=4$ and the number of nonzero rows of $\mathbf{X}$ was 18. The inter-vector correlation values were chosen from
the range -0.99 to 0.99, and the experiment was repeated for each of the values. The T-MSBL algorithm \cite{zhang2011IEEE}, a member of the T-SBL family introduced in Section \ref{subsec:inter}, was carried out to show the benefit from exploiting the correlation. For comparison, two typical MMV algorithms which do not exploit the correlation, namely M-SBL \cite{David2007IEEE} and Group-Lasso \cite{grouplasso} (the variant for the MMV model), were also performed. Note that if T-MSBL is forced not to exploit the inter-vector correlation (i.e., setting $\mathbf{B}_i=\mathbf{I}(\forall i)$), it reduces to the M-SBL algorithm.

The result (Fig.\ref{fig:Simulation12} (b)) shows that when the inter-vector correlation increases, T-MSBL has improved performance, but the two compared algorithms have degradation in performance.

\textbf{Experiment 3: Time-Varying Sparsity Model}. We conducted a noisy experiment to verify our strategy to treat a time-varying sparsity model as stated in Section \ref{subsec:vary}. $\mathbf{\Phi}$ was of the size $60 \times 256$. The column number of $\mathbf{X}$ was 50. The number of nonzero rows, $K$, during the first 15 columns of $\mathbf{X}$ was 15. $K$ was increased by 10 starting from the 16-th to the 31-th column of $\mathbf{X}$. Also, starting from the 26-th column, 5 existing nonzero rows were set to zeros. Each nonzero row was modeled as an AR(1) process with the AR coefficient varying from 0.7 to 0.99, and had a duration of at most 20 columns. SNR was 20 dB.

\begin{figure}[t]
\centering
\includegraphics[width=6cm,height=4.5cm]{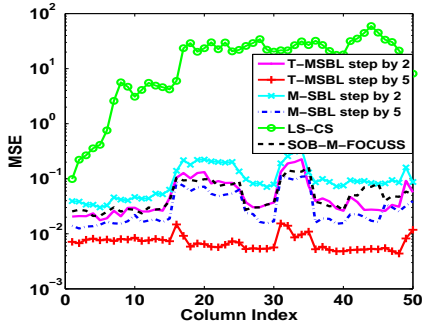}
\caption{Performance comparison in the experiment with time-varying sparsity.}
\label{fig:DCS}
\end{figure}

T-MSBL, M-SBL, SOB-M-FOCUSS, and LS-CS were compared. SOB-M-FOCUSS treats a time-varying sparsity model as a series of overlapped MMV models and exploits smoothness in amplitudes of non-zero entries of $\mathbf{x}_t$ over a short interval. For this algorithm, we set the length of each MMV model to 5, and set the overlapping rate to 0.5. Its smoothing matrix was a second-order smoothing matrix given in \cite{zhang2011IEEE}. LS-CS is an algorithm which does not exploit the benefit of multiple measurement vectors and the inter-vector correlation. SOB-M-FOCUSS and LS-CS were given the true noise variance, while both T-MSBL and M-SBL learned the noise variance. When performing T-MSBL and M-SBL, we approximated the time-varying sparsity model in two ways. One  was using the concatenation of 25 MMV models with each MMV model containing 2 columns. The second was using 10 MMV models with each containing 5 columns. Figure \ref{fig:DCS} shows the advantages of exploiting multiple measurement vectors (by comparing T-MSBL/M-SBL/SOB-M-FOCUSS to LS-CS) and of exploiting the inter-vector correlation (by comparing T-MSBL to M-SBL) by adaptively learning the correlation (by comparing T-MSBL to SOB-M-FOCUSS).

\section{Conclusion}
This paper discussed the problem of sparse signal recovery when there is correlation in the values of the non-zero entries. We reviewed both intra-vector correlation in the context of the block sparse model and intra-vector correlation in the context of the multiple measurement vector model. We discussed
how the sparse Bayesian learning framework can effectively incorporate correlation at the algorithm level. The impact of correlation on the limits of support recovery is also discussed. Since applications
involving sparsity are likely to be endowed with additional structure,
incorporating structure motivated by  applications and exploiting them to develop algorithms as well as to improve recovery performance
holds much promise.

\section*{Acknowledgment}
The work was supported by NSF grant CCF-0830612 and CCF-1144258.

\bibliographystyle{IEEEtran}
\footnotesize
\bibliography{zhilin,zhilin_sparse,MMVstrings,MMVmanuals}

\end{document}